\documentclass[tighten,iop,,appendixfloats]{emulateapj}
\usepackage{enumerate}

\begin{document}

\title{Quasar Cartography: from Black Hole to Broad Line Region Scales}
\author{Doron Chelouche\altaffilmark{1} and Shay Zucker\altaffilmark{2}}
\altaffiltext{1} {Department of Physics, Faculty of Natural Sciences, University of Haifa, Haifa 31905, Israel; doron@sci.haifa.ac.il}
\altaffiltext{2} {Department of Geophysical, Atmospheric, and Planetary Sciences, Raymond and Beverly Sackler Faculty of Exact Sciences, Tel Aviv University, Tel Aviv 69978, Israel; shayz@post.tau.ac.il}
\shortauthors{Chelouche \& Zucker}
\shorttitle{Quasar Cartography}

\begin{abstract}

A generalized approach to reverberation mapping (RM) is presented, which is applicable to broad- and narrow-band photometric data, as well as to spectroscopic observations. It is based on multivariate correlation analysis techniques and, in its present implementation, is able to identify reverberating signals across the accretion disk and the broad line region (BLR) of active galactic nuclei (AGN). Statistical tests are defined to assess the significance of time-delay measurements using this approach, and the limitations of the adopted formalism are discussed. It is shown how additional constraints on some of the parameters of the problem may be incorporated into the analysis thereby leading to improved results. When applied to a sample of 14 Seyfert 1 galaxies having good-quality high-cadence photometric data, accretion disk scales and BLR sizes are simultaneously determined, on a case-by-case basis, in most objects. The BLR scales deduced here are in good agreement with the findings of independent  spectroscopic RM campaigns. Implications for the photometric RM of AGN interiors in the era of large surveys are discussed.

\end{abstract}


\section{Introduction}

Active galactic nuclei (AGN) are thought to be powered by the accretion of material onto a supermassive black hole \citep{sal64,zel64}. In the standard paradigm of accretion disks the infalling gas is able to gradually lose angular momentum by viscous processes, as it makes its way to tighter orbits around the black hole, ultimately converting a fair fraction of its rest energy to radiation  \citep{ss73}.  In this picture, shorter wavelength photons are emitted by hotter gas inhabiting more compact regions around the black hole. Some indication for the relevance of this simple physical picture to quasars is given by reverberation mapping (RM) studies. In a nut shell, if perturbations in the inner accretion flow result in enhanced short-wavelength emission, which impinges upon the the outer accretion flow, then enhanced emission from larger scales would be a delayed (perhaps smeared) version of the short-wavelength light curve. Indeed, this scenario seems to be in excellent agreement with the observations of NGC\,7469, being the only object for which multi-epoch high-cadence optical spectroscopy revealed a statistically robust wavelength-dependent time-delay between continuum emission at different wavelengths \citep{col98}. Attempts to expand such studies to other AGN using photometric means revealed marginal signals, and a large scatter in properties, to which simple models provide unsatisfactory explanation \citep{ser05,cak07}.

Beyond the accretion disk lies the broad line region (BLR), which is responsible for the broad, typically a few $\times 10^3\,{\rm km~s^{-1}}$ wide, emission lines seen in the spectra of type-I AGN. Such lines respond to the flux variations of the continuum source, indicating that photoionzation is a key process setting their properties. While there is considerable uncertainty in the geometry and the physical origin of the BLR, this component of the active nucleus provides a valuable means for estimating the mass of supermassive black holes (SMBH), which power all quasars. Specifically, upon measuring the size of the BLR using RM \citep[and references therein]{kas00,ben09,den10}, and given a rather generic set of assumptions concerning its  kinematics, the obtained black hole masses are in agreement with other independent methods, as applied to nearby objects \citep{on04,woo10}. 

In its simplest form, RM of AGN seeks a single time scale, a lag, between two light curves that arise from physically distinct, yet causally connected, regions. For example, in BLR studies, pure continuum and pure emission line light curves are obtained by means of spectral decomposition, and various cross-correlation techniques may be used to determine the time lag, $\tau_l^{\rm spec.}$, between them \citep[and references therein]{pet04}. Nevertheless, properly distinguishing between emission which arises from two distinct regions in the AGN is not always possible. For example, line and continuum emission contribution to the signal are difficult to disentangle in regions of the spectrum where poorly resolved (e.g., iron) line blends are present. Also, dealing with photometric data, it is generally impossible to separate line from continuum emission hence standard cross-correlation analyses techniques may lead to erroneous results.

With the advance in large sky-coverage astronomy, the field is soon to be overwhelmed by high-quality photometric data for numerous quasars, in several broad wavelength bands\footnote{Several current and future surveys include OGLE, PanSTAARS, Gaia, as well as the {\it Large Synoptic Survey Telescope} (LSST).}. Clearly, harnessing the power of photometric surveys to probe the physics of the innermost regions of AGN is of importance, as it may shed light on long-standing questions concerning SMBH demography and accretion physics. Indeed, there have been several attempts to carry out continuum RM of the accretion disk in quasars using photometric means \citep{col01,ser05,bac09,kop10}. In addition, several recent studies have shown that RM of the BLR can also be carried out using narrow-band \citep{ha11,po12,po13} as well as broadband data \citep{cd11,cd12,ed12,po13}. Nevertheless, the degree to which reliable (non-degenerate) information about BLR {\it and} accretion physics may be obtained by pure photometric means, has not been explored.

In this paper we investigate the ability of photometric surveys to shed light on accretion disk and BLR physics in AGN. The paper is outlined as follows: in \S2 we discuss the limitations of standard cross-correlation analysis, and outline a new RM scheme in \S3 where we also critically assess its applicability by means of simulations. Section 4 applies the approach to publicly available data from \citet{ser05}. Some physical implications of our results are detailed in \S5 with a summary following in \S6. 

\section{Motivation}

\begin{figure}
\plotone{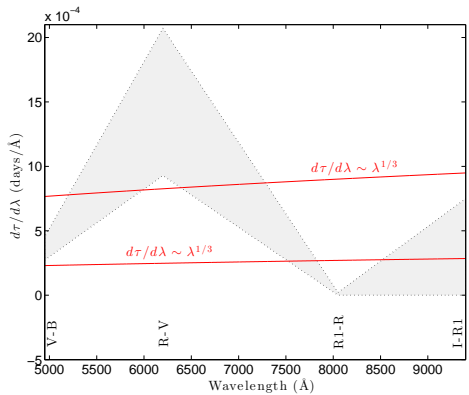}
\caption{The interband lag gradient after \citet[based on data from \citet{ser05}]{cak07}. Shown in gray is the range of values betwen the 30th to 70th percentiles (excluding 1E\,0754.6 and note that measurement uncertainties were not considered when evaluating $d\tau/d\lambda$). Predictions for $d\tau/d\lambda$ for the propagation of perturbations across a thin self-similar standard accretion disk, for which $\tau\sim\lambda^{4/3}$, are shown as red curves for two arbitrary normalizations. Clearly, the model does not trace well the data. In particular, the $R-V$ results seem to considerably deviate from naive model expectations, resulting in a non-monotonic behavior, which is due significant contribution of the BLR (H$\alpha$) to the signal in the $R$-band (see text).}
\label{cak}
\end{figure}

We have previously shown that line-to-continuum time-delays may be constrained using pure broadband photometric means \citep[and see also \citet{po13}]{cd11,cd12,ed12}. Conversely, this raises the concern that the contribution of emission lines to the broadband flux might influence time-delay measurements between adjacent continuum bands using photometric data \citep{col01,ser05,bac09}. That such a problem exists is already hinted by the results of \citet[see their Fig. 6]{wan97} and \citet[see their Fig. 6]{col98} who correlated continuum emission at different wavebands, and found that the wavelength-dependent lag, $\tau(\lambda)$, is considerably larger at wavelengths where line emission is present (but does not reflect on the line-to-continuum time-delay). 

There are good indications that interband time-delays, using broadband photometric data, suffer from similar effects, as can be seen in  in figure \ref{cak} after \citet{cak07}: the typical time-lag derivative $d\tau/d\lambda$ shows a jump around 6000\AA, which is likely driven by the fact that the $R$ band includes substantial contribution of the H$\alpha$ line to its flux \citep[see their Fig. 11]{cd11}. In contrast, simple models for irradiated accretion disks, which are often used to interpret the data \citep{ser05,cak07}, predict a monotonic function for $d\tau/d\lambda$ (Fig. \ref{cak}).

\begin{figure}
\plotone{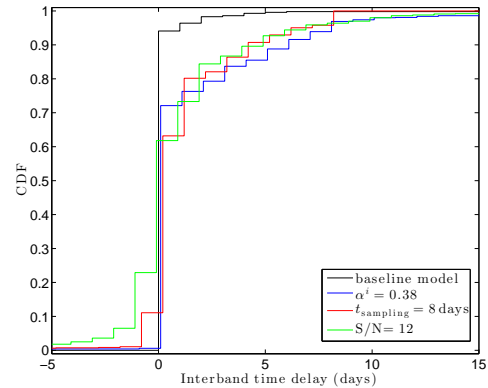}
\caption{Cumulative distribution functions (CDFs) for the measured interband time-delays, $\tau_c^{\rm 1D}$, for models with $\tau_c^i=0$\,days and $\tau_l^i=50$\,days. The baseline model is defined by daily sampling ($\tau_{\rm sampling}=1$\,day) over 50\,days, and having S/N$=100$, and $\alpha^i=0.23$. Excursions from the baseline model, changing one parameter at a time, correspond to different, slightly shifted for clarity, colored CDFs (see legend). All CDFs imply a finite mean interband time-delay although $\tau_c^i=0$\,days was assumed. Therefore, attempting to recover $\tau_c^i$ via interband time-delays may lead to erroneous results and unphysical constraints on accretion disk models.}
\label{biases}
\end{figure}

To gain better understanding of the effect that broad emission lines may have on the measured interband time-delay we resort to simulations akin to those used by \citet{cd11}: we model the continuum light curve, $f_c$, as a Fourier sum of independent modes with random phases whose amplitude, $A(\omega)\sim (1+\delta)\omega^{-\gamma/2}$, where $\gamma=2$, and $\delta$ is a random Gaussian variable with a standard deviation of 0.2 and a zero mean \citep{giv99}. We then construct a second broadband lightcurve, $f_{lc}$, which includes the contribution from continuum and line processes, both of which are delayed with respect to $f_c$, so that $f_{lc}=[(1-\alpha^i) \psi_c(\tau_c^i)+\alpha^i\psi_l(\tau_l^i)] *f_c$ ($'*'$ denotes convolution). Here, $\alpha^i$ is the broad emission line contribution to the broadband flux, and $\psi_c(\tau_c^i)$ and $\psi_l(\tau_l^i)$ are the (poorly constrained) continuum and line transfer functions, respectively, with $\tau_c^i,~\tau_l^i$ being their respective centroids. For simplicity, we consider the observationally-motivated thick shell BLR model of \citet[see their Fig. 2 for definition and \citet{mao91} for further details]{cd11} for both transfer functions and note that the results are not very sensitive to this choice. 

\begin{figure*}
\plottwo{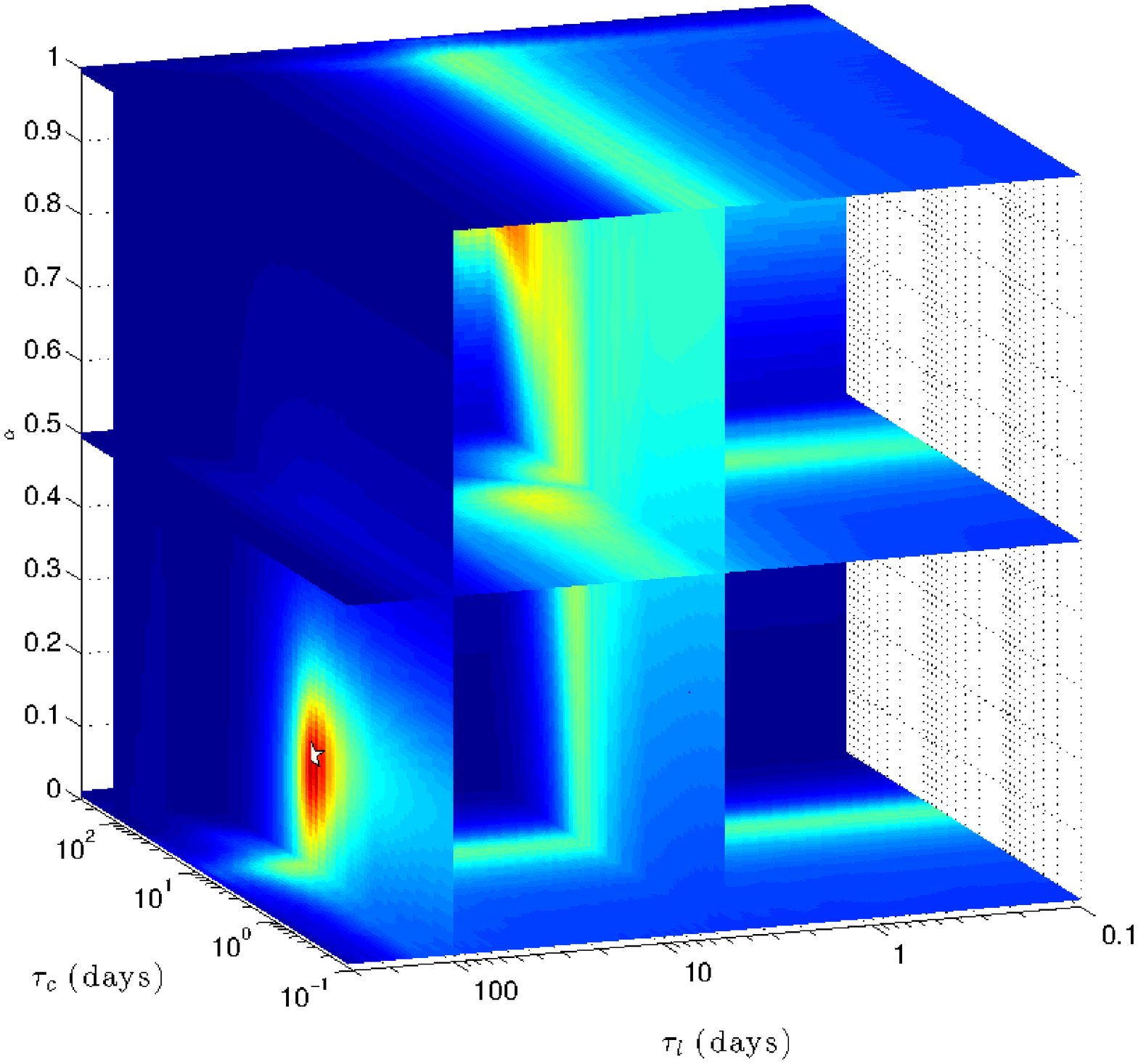}{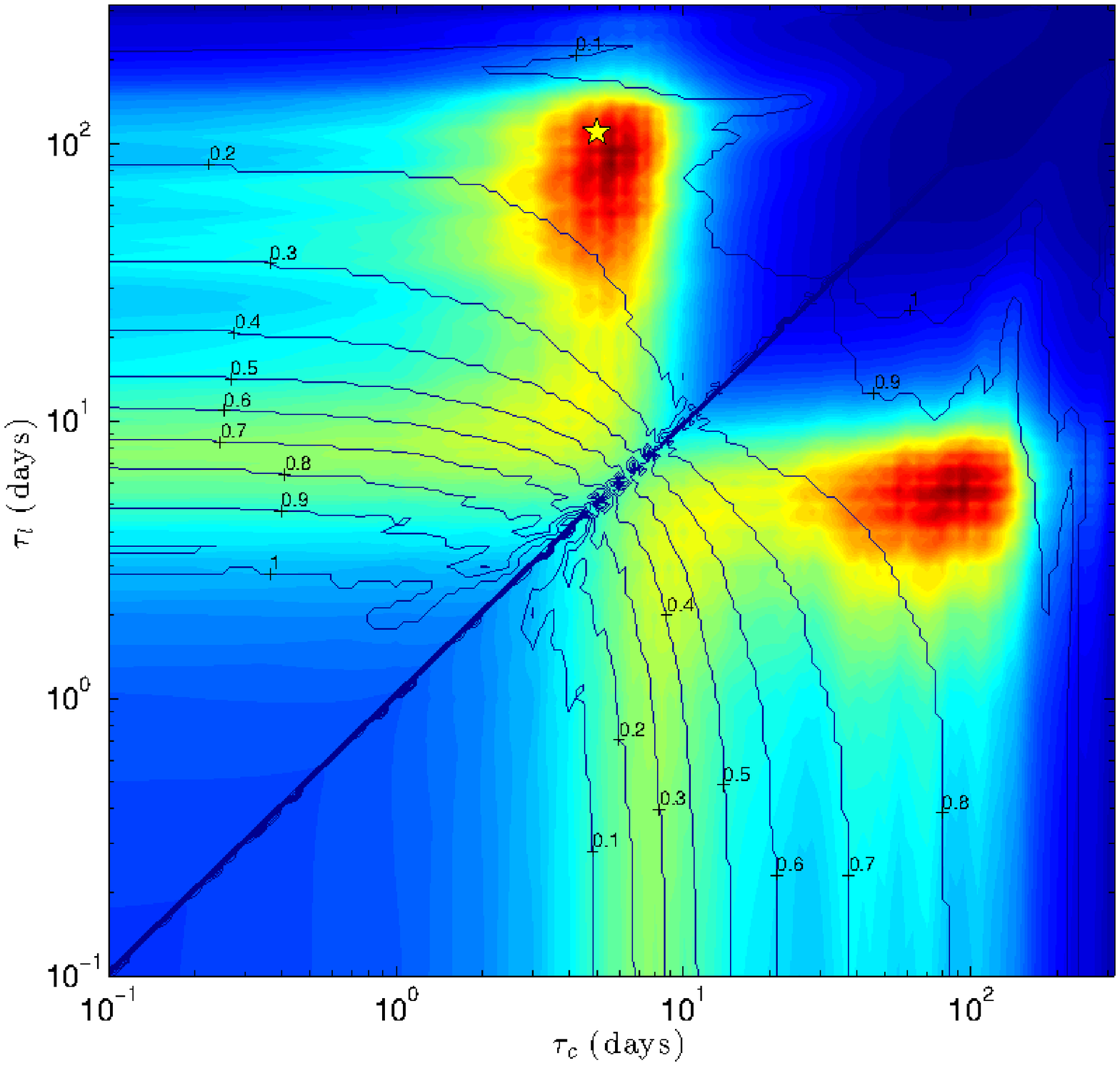}
\caption{The correlation function, $R(\tau_c,\tau_l,\alpha)$, as computed for our standard model ($\tau_c^i=5$\,days, $\tau_l^i=100$\,days, and $\alpha^i=0.17$). {\it Left:} a few two-dimensional (2D) slices in the three-dimensional (3D) parameter space are shown. Red-to-blue correspond to high-to-low values of the correlation function. Yellow pentagram shows the values of the input model . {\it Right:} 2D projection of the 3D correlation function. Values of $R$ in the 2D plane correspond to the maximum $R(\tau_c,\tau_l)$ along the third ($\alpha)$ dimension. Overlaid are contours showing $\alpha(\tau_c,\tau_l)$, which correspond to the value of $\alpha$ for which $R(\tau_c,\tau_l,\alpha(\tau_c,\tau_l))$ is maximized. Clearly, the correlation function is symmetric with respect to the diagonal (small asymmetries have to do with grid interpolations). This example shows that the proposed formalism is able to recover the input model values (see text).}
\label{3D}
\end{figure*}

For the set of models considered in this section we take $\tau_c^i=0$\,days and a finite $\tau_l^i$, and look for interband time-delays for a range of observing and sampling conditions. As expected, a lag-distribution is obtained on accounts of the different light curve realizations (Fig. \ref{biases}). The particulars of the distributions depends on the emission line properties, the sampling, and the signal-to-noise (S/N). Nevertheless, the mean of those distributions is generally biased to positive lags implying that interband time-delays cannot be used to reliably constrain accretion disk physics. The bias increases with the fractional contribution of the emission line to the band, and is larger for poorer sampling. The effect of the number of points in the light curve, as well as the presence or absence of light curve gaps (so long as adequate sampling is maintained) is, generally, secondary. Lower signal-to-noise (S/N) observations result in a larger dispersion of the measured lags yet does not wash away the bias.

In light of the above analysis, one must consider the possibility that some of the measured interband time-delays, in any sample \citep{ser05,bac09}, have little to do with accretion disk physics but rather indirectly reflect on the broad emission line "contamination" of one or more of the bands. The problem may be further aggravated should additional slowly varying continuum components be present, such as hot dust emission close to the nuclear source \citep{gas07}. The limitation of current analysis methods, and the direction headed by future photometric surveys, warrant further investigation into improved analyses techniques that may be able to alleviate the degeneracy between line- and continuum-induced delays.

\section{Method}

Consider a simplified scenario wherein a quasar is observed in two wavelength bands leading to two light curves: $f_c(t)$ to which only continuum emission processes contribute, and $f_{lc}(t)$ to which both slightly lagging continuum emission, and an additional, slowly varying and lagging component contribute. Without loss of generality, we shall assume the latter component is associated with broad line emission. 

As previously noted, the details of the transfer functions, $\psi_c$ and $\psi_l$ are poorly known. Nevertheless, to zeroth order approximation one may write\footnote{We assume $f_c$ and $f_{lc}$ are the fluxed versions of the lightcurves.},
\begin{equation}
f_{lc}(t)\simeq f^m_{lc}(t;\tau_c,\tau_l,\alpha)\equiv (1-\alpha) f_c(t-\tau_c) +\alpha f_c(t-\tau_l), 
\label{flcp}
\end{equation}
where $\alpha$ ($0\le \alpha \le 1$) is the contribution of the broad emission line to the broadband flux. Observationally, neither the time lags, nor $\alpha$ are known, although the latter may often be estimated if, for example, a single epoch spectrum (or some general knowledge of quasar spectra) is available and the broadband filter transmission curve is known \citep[and references therein]{cd11}. Using the above approximation to $f_{lc}$ we are neglecting higher order moments of the transfer functions other than their centroids. As we shall show below, this approximation is often adequate for recovering the lags.

There are several possible ways to deduce the set of parameters $[\tau_c,\tau_l,\alpha]$ for which $f^m_{lc}$ provides the best approximation to $f_{lc}$. Here we use standard correlation analysis, and define the multivariate correlation function (MCF),
\begin{equation}
R(\tau_c,\tau_l,\alpha)=\frac{\sum_i [f_{lc}(t_i)-\bar{f}_{lc}] [f^m_{lc}(t_i;\tau_c,\tau_l) -\bar{f}^m_{lc}(\tau_c,\tau_l)]}{N\sigma_{f_{lc}} \sigma_{f^m_{lc}}(\tau_c,\tau_l)},
\label{ccf}
\end{equation}
where barred quantities are averages, and $\sigma$ denote standard deviations. We emphasize that all quantities on the righthand side of equation \ref{ccf} depend, either explicitly or implicitly, on $\tau_c$ and $\tau_l$ as only overlapping portions of the light curves are being evaluated \citep{wel99}. The interpolation method used here to compute equation \ref{ccf} for general light curves is outlined in the Appendix. To constrain $\tau_c,~\tau_l$ and $\alpha$, we seek points in the three-dimensional (3D) parameter space, defined by $(\tau_c,~\tau_l,~\alpha)$, which maximize $R$. 

Two interesting limits of the above correlation function are the following: when pure line and continuum light curves are available (e.g., via spectral decomposition), $\alpha^i=1$, and the formalism converges to the standard cross-correlation technique employed by RM studies. When broadband data are concerned so that $\alpha^i \ll 1$, and continuum transfer effects are negligible (i.e., when setting $\tau_c=0$\,days), it can be easily shown that the formalism converges to the scheme adopted by \citet{cd11}\footnote{See  the Appendix of \citet{zm94} and note that, using their notation, and Taylor expanding in the limit $\alpha \ll 1$, while setting $s_1=0$, results in the expression defined by \citet{cd11} up to an additive constant and an overall scaling factor.}.

As the above formalism is not restricted to a particular value of $\alpha$, the method is applicable also to narrow-band data, for which, typically, $\alpha \lesssim 1$. Unlike recent implementations of narrow-band RM that seek a maximum in the cross-correlation of $f_c$ and $f_l^m$, where $f_l^m\equiv f_{lc}-\alpha f_c$, and a spectrally-motivated value for $\alpha$ is assumed \citep[and references therein]{po13}, in our formalism, no prior knowledge of $\alpha$ is required, and its value is being constrained by the requirement for a maximal $R$-value within the computational domain. The proposed method can also be applied to spectroscopic data sets where it can alleviate the need for spectral decomposition of line and continuum signals. This can be especially beneficiary in regions of the spectrum where spectral decomposition into continuum and line features is challenging, such as near iron emission line blends \citep{raf13}.

In what follows we apply the method for broadband photometric data.

\subsection{Solutions for $\tau_c,~\tau_l$ and $\alpha$}

To demonstrate the ability of the above formalism to constrain the relevant model parameters using broadband photometric data, we resort to simulations of the type described in \S2. For the particular case shown in figure \ref{3D}, we assume $\tau_c^i=5$\,days, $\tau_l^i=100$\,days, and $\alpha^i=0.17$. 500 daily visits are assumed in each band, and a measurement uncertainty of 1\% is considered. We show several slices in 3D space where red colors correspond to high $R$-values. Two solutions are evident where $R$ peaks: $[\tau_c,\tau_l,\alpha]=[5,100,0.17]$ and $[\tau_c,\tau_l,\alpha]=[100,5,0.83]$. That these two solutions are in fact the same is evident from the definition of $f^m_{lc}$, which is symmetric with respect to $\tau_c \rightleftharpoons \tau_l$ and $\alpha \rightleftharpoons 1-\alpha$ interchanges, and results from the fact that both the continuum and emission line templates, are identical\footnote{This is different than the case considered by \citet{zm94} in their search for spectroscopic binaries, where different templates were used to reconstruct the combined spectrum of the system.}.

We note that the dependence of $R$ on $\alpha$ is relatively simple in the sense that, for a particular choice of $[\tau_c,\tau_l]$,   one maximum will be obtained along the $\alpha$-dimension \citep[see their Eq. A2 and its following derivatives]{zm94}. This allows us to reduce the general problem to that of finding the maximum of $R$ in the 2D plane where $R=R(\tau_c,\tau_l,\alpha(\tau_c,\tau_l))$, and where $\alpha(\tau_c,\tau_l)$ is the value of $\alpha$ which maximizes $R$ given $\tau_c$ and $\tau_l$. The 2D projection is also shown in figure \ref{3D}, over-plotted with contours of $\alpha(\tau_c,\tau_l)$. The maxima described above are clearly evident, as well as the symmetric nature of the correlation function with respect to the diagonal. 

Besides the prominent peak, for which the deduced parameters satisfy $[\tau_c,\tau_l,\alpha]\simeq[\tau_c^i,\tau_l^i,\alpha^i]$, we note a ridge extending from the peak down just above the line where $\tau_c/\tau_l\sim 1$ with $\alpha\sim 0.5$. The fact that $R$ is relatively large over this ridge, although {\it not} at maximum in the 2D plane, has to do with the fact that the somewhat broadened and lagging shape of $f_{lc}$ with respect to $f_c$ may be (poorly) reconstructed by two, slightly offset in time, versions of $f_c$, with comparable weights. 

To be able to separately deduce $\tau_c$ and $\tau_l$ depends on our ability to disentangle the two sources contributing to the {\it total} transfer function giving rise to $f_{lc}$, namely $\psi_{lc}=(1-\alpha^i)\psi_c+\alpha^i\psi_l$. Clearly, there are situations where this will not be possible: for example, if $\alpha^i \lll 1$ then, under realistic noisy observing conditions, it will not be possible to detect line emission, and a solution, which is insensitive to $\tau_l^i$ will be obtained. At the other extreme, if $\alpha^i \to 1$ then no continuum component will be observed, and the solution will be insensitive to $\tau_c^i$. Nevertheless, because in those two limits, $f_{lc}$ is essentially a somewhat broadened and shifted version of $f_c$ (on accounts of $\psi_c$ or $\psi_l$), a superior fit to the light curve may be obtained by a linear combination of roughly equally proportioned $f_c$ templates slightly shifted with respect to the relevant lag. For example, for $\alpha^i \to 1$, a solution may be obtained with $\alpha\sim 0.5$ and $\tau_c \lesssim \tau_l^i$ and  $\tau_l \gtrsim \tau_l^i$ so that neither of the deduced lags reflects directly on the physics of the BLR, yet their average does. Similarly, when $\alpha^i \to 0$, $\vert \tau_c\vert \sim \vert\tau_l\vert\sim \tau_c^i$ solutions will be obtained with $\alpha\sim 0.5$. Clearly, in such limiting cases, the physics of only one region may be constrained, and a more straightforward model to consider is by setting $\alpha=0$ or $\alpha=1$, in which case a simple cross-correlation scheme is recovered.

\begin{figure}
\plotone{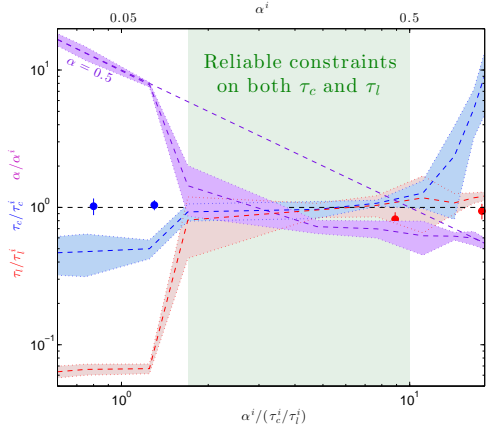}
\caption{Deviations of the recovered parameter values from their input values as a function of $\alpha^i/(\tau_c^i/\tau_l^i)$ (bottom axis), or $\alpha^i$ (top axis), for our standard model (see text) but with a varying $\alpha^i$ (all variables are solved for, and no priors are assumed). Blue (red and green) dashed curve shows the mean $\tau_c/\tau_c^i$ ($\tau_l/\tau_l^i$ and $\alpha/\alpha^i$) for the ensemble, and one standard deviation around the mean. Clearly, the input values may be  recovered for $\tau_c^i/\tau_l^i\lesssim \alpha^i\lesssim0.5$, with $\alpha\lesssim \alpha^i$ due to our neglect of higher moments in $\psi_l$ (see text). Outside this range, $\alpha \sim 0.5$ is obtained, and either $\tau_c^i$ or $\tau_l^i$ may be recovered by averaging over the lag results (see text and the red and blue points with error-bars).}
\label{check}
\end{figure}

To better understand the limitations of our formalism, we carried out sets of simulations wherein quasar light curves were constructed given $[\tau_c^i,\tau_l^i,\alpha^i]$, and the deduced $[\tau_c,\tau_l,\alpha]$  from the 3D correlation function logged. Comparing the input values to the deduced ones, we find the following trends: it is possible to reliably deduce the input lags so long as $\tau_c^i/\tau_l^i\lesssim \alpha^i\lesssim0.5$. In this case, $\alpha\lesssim \alpha^i$, which results from the fact that $f_{lc}^m$ neglects higher moments of the line transfer function other than its centroid.  For $\alpha^i$ outside this range, $\alpha\sim 0.5$ is obtained, and two limits may be defined: the pure continuum limit, and the pure line limit. The first limit applies when $\alpha^i \ll \tau_c^i/\tau_l^i$, and the continuum-to-continuum time-delay is given by $(\tau_c+\tau_l)/2$, and no reliable information may be obtained concerning line emission in this case. In the line-dominated limit, the line-to-continuum time-delay is obtained using a similar expression and the data cannot be used to determine continuum time-delays. Therefore, caution is advised when interpreting cases for which $\alpha\sim 0.5$ is obtained. The above is graphically summarized in figure \ref{check} for a particular set of simulations with $\tau_c=5$\,days and $\tau_l=100$\,days.

In case the identity of $f_c$ and $f_{lc}$ is unknown (as would be the case if, for example, the contribution of varying emission lines to the respective bands is unknown), or if one wishes to keep an open mind concerning the propagation of perturbations across the accretion flow, one should extend the 3D correlation analysis to negative values of $\alpha$ and $\tau_c$. Specifically, by reversing the choice of light curves ($f_c \rightleftharpoons f_{lc}$), the solution transforms such that $[\tau_c,\tau_l,\alpha] \to [-\tau_c,\tau_l,-\alpha] $. Deviations from simple solution symmetry will occur when $\tau_c^i/\tau_l^i \lesssim 1$, which might be of some relevance to highly ionized broad emission lines. 

\subsection{Error estimation, significance, and the use of priors}

The common practice in the field of RM is to use the FR-RSS scheme for estimating the uncertainty on the deduced lags \citep{pet98}. Briefly, the flux randomization (FR) part of the algorithm accounts for the effect of measurement uncertainty, while the purpose of the random subset selection (RSS) scheme is to check the sensitivity of the result to sampling by randomly selecting sub-samples of the data and operating on those \citep{pet98,cd11}. As such, the FR-RSS algorithm provides a combination of error and significance estimates. 

Although the RSS algorithm is mathematically well-defined, it is, in fact, arbitrary, as it discards, on average, a certain fraction of all visits, often leading to over-estimated errors (S. Kaspi, private communication). An additional shortcoming of the RSS approach is that it transforms, by construction, an evenly-sampled time-series, to an unevenly sampled one, with all the related complications \citep{ed12}. This means that, although useful in some cases, the RSS scheme may not be generally applicable. In particular, when applied to our problem, the RSS scheme tends to unjustifiably suppress the signal and mix two physically-distinct solutions (see below), leading to measurement uncertainties that are driven more by systematic effects than statistical ones. Lastly, using the FR-RSS scheme, it is not clear how to estimate the detection significance of an emission line signal lurking in the data; recall, that broadband light curves are to zeroth order identical, and that an emission line signal may not be easily discernible by eye.

The approach taken here is different: we separately treat the question of emission line lag uncertainty and its significance\footnote{It is beyond the scope of this paper to provide a general purpose error and significance estimation algorithm, which would be applicable to all RM studies, and note that a potentially promising route may involve light curve modeling via Gaussian processes \citep{rp92,zu11}}. We estimate the uncertainty on the deduced lags and $\alpha$ using the FR part of the FR-RSS algorithm. More specifically, if $f_c(t)~[f_{lc}(t)]$ is characterized by measurement uncertainties $\sigma^e_c(t)~[\sigma^e_{lc}(t)]$, then new light curves may be reconstructed from the original data so that: $f_c(t) \to f_c(t)+\delta(\sigma^e_c(t))$ [and similarly $f_{lc}(t) \to f_{lc}(t)+\delta(\sigma^e_{lc}(t))$], where $\delta$ is a random Gaussian variable with a zero mean and a standard deviation being the measurement uncertainty. The 3D cross-correlation analysis is repeated for many light curve realizations thereby generating 3D correlation peak distribution, allowing us to estimate  the uncertainty on each of the parameters\footnote{As this scheme does not remove points from the light curves during error-estimation, it could benefit from initial screening of the light curves against bad data. With upcoming surveys, having robust data quality checks and uniform reduction and calibration schemes, we expect bad data issues to become less relevant.}.

\begin{figure}
\plotone{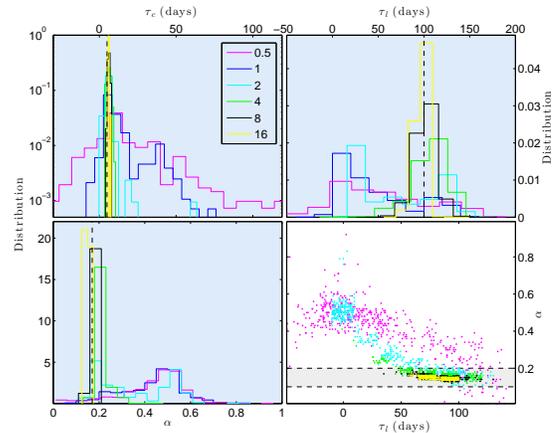}
\caption{Error-estimation on the deduced quantities and the effect of measurement noise. {\it Top-left:} shows the $\tau_c$ distributions for our standard model, and for varying levels of noise denoted by $\alpha^i/(\sigma^e/\bar{f})$ (see legend). As expected, larger noise results in broader non-Gaussian distributions. {\it Top-right:} the $\tau_l$ distributions peak at $\tau^i_l$ (dashed lines) for large values of $\alpha^i/(\sigma^e/\bar{f})$, but develops a two peak form for smaller values (see text). Likewise is the case for the $\alpha$-distribution (bottom-left panel). {\it Bottom-right:} show the interdependence between the values of the deduced $\alpha$ and $\tau_l$. The use of the prior $0.1<\alpha<0.2$ can be used to constrain $\tau_l$ even in relatively noisy data (see text).}
\label{fr}
\end{figure}

Figure \ref{fr} shows the time lags and $\alpha$ distribution functions for the model considered in figure \ref{3D}, and for several levels of measurement noise. As expected, the best constrained parameter at any noise level is $\tau_c$ due to the dominance of continuum emission processes in quasar light curves. Results for both $\tau_c$ and $\tau_l$ are most reliable for $\alpha > \sigma^e/\bar{f}$ where $\sigma^e$ is the typical noise level in the bands, and $\bar{f}$ the mean flux level. For larger values of $\sigma^e/\bar{f}$, the lag distribution functions significantly broaden, are clearly non-Gaussian, yet their modes are around the input lags. When $\sigma^e/\bar{f}\sim \alpha^i$, the emission line lag distribution may become double peaked, with one peak around $\tau_l^i$, and the second peak at shorter times, of order $\tau_c^i$. This behavior results from the emission line signal being gradually washed out by noise, and equally favorable agreement with the data is obtained by ignoring its contribution. Under such conditions, a double humped $\alpha$-distribution appears for $\sigma^e/\bar{f}\sim \alpha^i$, with one peak at around $\alpha^i$, and the other at $\alpha\sim 0.5$ (in our experience, an implementation of the FR-RSS scheme to such cases unjustifiably degrades the signal, and leads to poorly resolved solutions). For still larger values of $\sigma^e/\bar{f}$, the line signal is quenched and only $\alpha\sim 0.5$ and $\tau_c\sim \tau_l\sim \tau_c^i$ peaks persist, as discussed above. In what follows, and unless otherwise specified, we identify the parameter values with the more pronounced peaks of their respective distributions, with an uncertainty interval encompassing 68\% on either side of the peaks\footnote{This ensures that the peak identified does not fall outside the percentile intervals, as may occur for highly skewed, or double peaked, distributions.}.

It is important to realize that there may be inter-dependencies among the deduced values of the parameters. Specifically, at high levels of measurement noise, a large range of $\tau_l$ and $\alpha$ values is consistent with the data, as can be seen from their broadened distributions, yet larger values of $\alpha$ also correspond to solutions with shorter $\tau_l$. Therefore, using prior information on the value of $\alpha$, if available, could help to constrain $\tau_l$ even in cases where $\sigma^e/\bar{f}\sim \alpha^i$. That this is the case is is shown in the bottom-right panel of figure \ref{fr}: at high noise levels both $\alpha$ and $\tau_l$ are significantly anti-correlated and span a large range of values with the most probable $\tau_l$ being at $\sim 0$\,days. However, by setting the constraint $0.1<\alpha<0.2$ (note the shaded region in Fig. \ref{fr} and recall that $\alpha^i=0.17$), the allowed $\tau_l$ range considerably narrows, and the physically relevant peak of the distribution function a identified at $\sim 100$\,days, i.e., consistent with the input value. 

\begin{figure}
\plotone{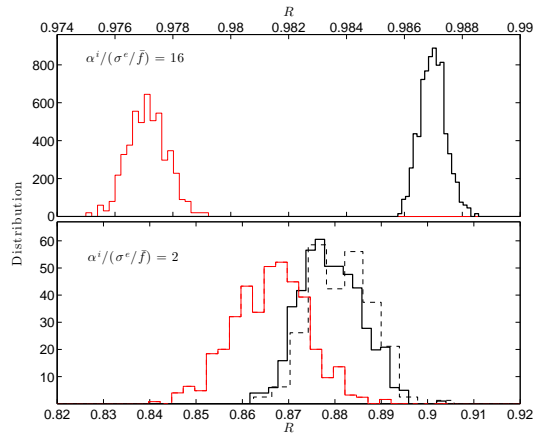}
\caption{Cross-correlation coefficient, $R$, distributions for the real and permutated (i.e., using $\tilde{f}_c$; see text) datasets for two levels of noise denoted in each panel. Black histograms correspond to the $R$-distribution for the real data, while red histograms are the outcome of the permutation scheme. Clearly, for the low noise levels characterizing our standard model, there is no overlap in the $R$-distributions between the real and permutated data implying a highly significant result. For higher noise levels (lower panel), there is considerable overlap and the probability that the mode of the $\tau_l$-distribution is, in fact spurious, is of order 10\% for the chosen example. Note also the lower $R$-coefficients obtained for noisier data. The effect of the prior $0.1<\alpha<0.2$ is also shown (dashed histogram), which renders the result highly significant as only $\alpha<0.1$ lead to $R>0.82$ for the permutated dataset (not shown). Clearly, the use of priors may significantly boost the significance of the results.}
\label{sig}
\end{figure}

\begin{table*}
\begin{center}
\caption{The \citet{ser05} sample of AGN}
{\tiny
\begin{tabular}{|lclllll|llllll|}
\tableline
&&&&&&&&&&&& \\
\multicolumn{7}{|c|}{Object Properties \& Sampling} & \multicolumn{6}{c|}{Photometric Reverberation Results} \\   
&&&&&&&&&&&& \\
\tableline
& & $\lambda L_\lambda $ & Light curve & $\tau_c^{\rm 1D}$ & $\tau_l^{\rm spec.}$  &  &  $\tau_{c}$ &  $\tau_{l}$ &  &  $\tau_{c}^{\rm prior}$ &  $\tau_{l}^{\rm prior}$ &   \\
Object ID  & $z$   & $[{\rm erg\,s}^{-1}]$ & properties & $[{\rm days}]$ & $[{\rm days}]$  & $\alpha^i$ & $[{\rm days}]$  & $[{\rm days}]$ & $\alpha$  & $[{\rm days}]$  & $[{\rm days}]$ & $P$ \\
(1) & (2) & (3) & (4) & (5) & (6) & (7) & (8) & (9) & (10) & (11) & (12) & (13) \\          
\tableline

1E\,0754.6+3928 & 0.096 &  $8$ & 3/80/2 & $8^{+4}_{-9}$  &  $72^{+265}_{-35}$ & $0.1^{({\rm a})}$ & $-2^{+10}_{-6}$ & $11^{+53}_{-3}$ & $0.18^{+0.43}_{-0.13}$ & $8^{+5}_{-10}$ & $64^{+11}_{-17}$ & $0.64$\\

3C\,390.3 & 0.056 & $2.4$ & 3/92/14 & $6.0^{+1.3}_{-3.6}$ & $28^{+8}_{-12}$ & 0.2  & $2^{+2}_{-1}$ & $120^{+20}_{-15}$ & $0.1\pm0.1$ & $2^{+2}_{-1}$ & $110^{+47}_{-20}$ & $0.58$ \\

Akn\,120 & 0.033 & $2.3$ &  3/75/8 & $2^{+0.8}_{-0.7}$ & $43^{+13}_{-11}$ & $0.2^{({\rm d})}$ & $1.5^{+1.0}_{-3.0}$ & $32^{+8}_{-24}$ & $0.1^{+0.33}_{-0.05}$ & $1.6^{+0.9}_{-2.0}$ & $16^{+9}_{-6}$ & $0.65$\\

MCG+08-11-011 & 0.020 & $1.4$ & 3/69/7 & $3.4^{+1.0}_{-0.9}$ & $30^{({\rm e})}$  & - & $2.5^{+0.8}_{-1}$ & $48^{+12}_{-4}$ & $0.09\pm0.03$ & $-$ & $-$ & $0.55$ \\

Mrk\,335 & 0.026 & $0.6$ & 3/66/8 & $2.5^{+0.7}_{-3.0}$ & $16\pm7$ & $0.17^{({\rm b})}$ & $-0.5^{+2.8}_{-2.3}$ & $3^{+2}_{-1}$ & $0.6^{+0.1}_{-0.2}$ & $1.5\pm1.0$ & $13^{+5}_{-2}$ & $0.97^{({\rm f})}$ \\

Mrk\,509 & 0.035 & $2.5$ & 3/58/5 & $6^{+3}_{-1}$&  $79^{+6}_{-7}$ & 0.3 & $1.6^{+0.8}_{-1.0}$ & $45^{+5}_{-15}$ & $0.18\pm0.04$ & $0.6^{+0.8}_{-0.2}$ & $30^{+15}_{-10}$ & $0.78$ \\

Mrk\,6 & 0.019& $1.6$ & 2/126/6 & $0.6^{+1.1}_{-0.9}$ & $10\pm1^{({\rm b})}$ & $0.17^{({\rm b})}$ & $-0.5\pm1.0$ & $5^{+4}_{-2}$ & $0.4^{+0.1}_{-0.2}$ & $0.6^{+1.0}_{-0.4}$ & $11^{+6}_{-5}$ & $0.85$ \\

Mrk\,79 & 0.022 &  $0.9$ & 2/73/9 & $2.5^{+1.6}_{-2.2}$ & $13^{+10}_{-8}$ & $0.16^{({\rm d})}$ & $0.4^{+0.1}_{-0.9}$ & $7^{+4}_{-1}$ & $0.3^{+0.2}_{-0.2}$ & $0.5^{+0.1}_{-0.2}$ & $11^{+23}_{-2}$ & $0.90$ \\

NGC\,3227 & 0.004 & $0.07$ & 2/102/2 & $0.6^{+0.8}_{-0.1}$ & $15^{+9}_{-11}$ & $0.15^{({\rm c})}$ & $0.5\pm0.25$ & $37^{+13}_{-10}$ & $0.10\pm0.05$ & $0.5^{+1.0}_{-0.3}$ & $47^{+6}_{-10}$ & $0.52$ \\

NGC\,3516 & 0.009& $0.4$ & 2/113/3 & $0.6^{+1.7}_{-1.3}$ & $10^{+7}_{-4}$ & 0.25 & $0.5\pm1$ & $23^{+2}_{-9}$ & $0.25\pm0.05$ & $0.5\pm1$ & $18^{+6}_{-4}$ & $0.99$ \\

NGC\,4051 & 0.002&  $0.01$ & 1/155/3 & $0.4^{+0.2}_{-1.0}$ & $6.0^{+2.5}_{-1.8}$ & 0.1 & $0.4^{+0.1}_{-0.8}$ & $20\pm10$ & $0.2\pm0.1$ & $0.4^{+0.1}_{-0.8}$ & $20^{+10}_{-4}$ & $0.80$\\

NGC\,4151 & 0.003 & $0.08$ & 2/149/10 & $0.6^{+0.2}_{-0.1}$ & $6^{+4}_{-3}$ & 0.2 & $0.3\pm0.1$ & $5^{+5}_{-1}$ & $0.2\pm 0.04$ & $0.3\pm0.2$ & $5\pm1$ &  $0.91$ \\

NGC\,5548 & 0.017 & $0.6$ & 2/141/7 & $7.5^{+2.0}_{-0.6}$& $15^{+5}_{-4}$ & 0.3  & $-0.5\pm1$ & $16^{+3}_{-4}$ & $0.40^{+0.10}_{-0.05}$ & $-0.5^{+1.0}_{-0.8}$ & $19^{+6}_{-2}$ & $0.99$ \\

NGC\,7469 & 0.016 & $0.7$ & 2/88/3 & $0.6^{+0.9}_{-0.1}$ & $4.8^{+1.5}_{-0.8}$ & 0.1 & $0.5^{+0.1}_{-0.2}$  & $19^{+3}_{-12}$ & $0.1\pm0.1$ & $0.5\pm0.1$ & $9^{+6}_{-9}$ & $0.60$ \\

\tableline

\end{tabular}
}
\end{center}
Properties and analysis results for the \citet{ser05} sample of photometrically monitored AGN. Columns: (1) Object name, (2) redshift, (3) monochromatic luminosity at 5100\AA\ in units of $10^{44}\,{\rm erg~s^{-1}}$ (see text), (4) Sampling properties [median sampling period/number of visits (the minimum between the $V$ and the $R$ bands)/the reduced variability measure in per cent (the minimum between the $V$ and the $R$ bands), as defined in \citet{kas00}], (5) interband time delays, as deduced from simple, 1D cross-correlation analysis (positive values indicate that the $R$-band lags the $V$-band), (6) Spectroscopically determined Balmer lines' time-delays [unless otherwise specified, taken from \citet{pet04} after averaging over all significant Balmer line results], (7) spectrally estimated broad emission line contribution to the $R$-band [using the spectral deconvolution method of \citet{cd11}, and the results of \citet[unless otherwise noted]{kas05}], (8)-(10) $\tau_c,~\tau_l$, and $\alpha$ values at which the cross-correlation coefficient, $R$, is maximized, (11)-(12) the deduced time-delays which maximize $R$ under the prior that $\alpha=\alpha^i$, (13) the confidence level of the solution ($1-P$ is the probability of the solution being due to chance occurrence and without invoking priors). \\
$^{({\rm a})}$Taken from \citet{ser07}. 
$^{({\rm b})}$Taken from \citet{gr12}. 
$^{({\rm c})}$Using data from \citet{sal94}.
$^{({\rm d})}$Using data from \citet{pet98a}. 
$^{({\rm e})}$No spectroscopic lag exists. Lag is roughly estimated from the $R-L$ relation of \citet{ben09}.
$^{({\rm f})}$Confidence level was calculated within the restricted (i.e., assuming priors) domain.
\end{table*}

A separate question to address concerns the significance with which one may claim to have detected an emission line signal lurking in the data (recall that the broadband light curves are, to zeroth order, identical). To this end, we use the following algorithm, which is applicable to cases where the line contribution to the broadband flux is small: we define $f_{lc}^m(\tau_c,\tau_l,\alpha)=(1-\alpha)f_c(\tau_c)+\alpha \tilde{f}_c(\tau_l)$ where $\tilde{f}_c$ is a {\it randomly-permutated} version of $f_c$, which preserves the number of points, and observing times, but shuffles the fluxes between the visits. This procedure is iterated many times, the 3D correlation function repeatedly evaluated, and its maximum value logged for each realization. The probability that our result is spurious is then given by  the probability for obtaining a correlation coefficient that is higher than that which is determined for the real (non-permutated) data.

Results for the model discussed above (see also Fig.\ref{3D}) are shown in figure \ref{sig}, and imply that, for low noise levels (i.e., $\alpha^i/(\sigma^e/\bar{f}) \gg 1$), a highly significant result is obtained as the real and the simulated (using the $\tilde{f_c}$ template) $R$-distributions do not overlap. For higher noise levels, the two distributions (using the real and shuffled data) begin to overlap, as can be seen for the case $\alpha^i/(\sigma^e/\bar{f}) =2$, and the probability that the result is spurious is non-negligible. 

We note that the use of priors can also boost the significance of the result. Specifically, our procedure for estimating the confidence level yields $R$-distribution functions that are similar to the observed one only for $\alpha<0.1$. When setting the $0.1<\alpha<0.2$ prior, the two distributions do not overlap, and the result is, in fact, highly significant. Therefore, the presence of finite $\alpha$ solutions provides a measure for the significance of the result.

\section{Application to broadband photometric data of low-luminosity AGN}

Here we implement our formalism to broadband photometric data of AGN. Specifically, we analyze the $V$ and $R$ broadband photometric light curves of 14 low-luminosity AGN from the \citet{ser05} sample; see table 1\footnote{This particular dataset has been previously used to shed light on accretion disk physics in AGN \citep{dor05,cak07,bre09}}. For this filter combination and the relevant redshift range, the $R$-band is relatively emission-line-rich due to the significant contribution of H$\alpha$ to its flux, while the $V$-band is line-poor\footnote{We note that for nearby objects and the filter scheme used by \citet{ser05}, the contribution of non-powerlaw (i.e., non-continuum, including line and line blends) emission to the $R$-band is at the $\sim 30\%$ level, while that which contributes to the $V$ band is at the $\sim 5$\% level, as the H$\beta$ line falls just shortward of the $V$-band.}. 

The optical luminosities quoted in table 1 follow the estimates given in \citet[taking the mean flux level in their Table 2]{cak07}, assuming concordance cosmology, their reddening corrections, and subtracting their best-fit value for the host contribution to the $V$-band flux. These luminosities may substantially differ from those quoted in other works [e.g., \citet{ben06} and \citet{ben09}] and may reflect on the actual variability of the sources, as well as on systematics in the luminosity determination affecting different works. As we are only interested in qualitative lag-luminosity relations, we do not quote luminosity uncertainties in this paper. For comparison purposes table 1 reports also on the interband time delay, $\tau_c^{\rm 1D}$, between the $V$ and $R$ bands. 

Our aim in this section is two-fold: 1) to constrain time-delays between continuum emission components that contribute to the two  bands, thereby assessing the reliability of simple interband cross-correlation techniques in shedding light on accretion disk physics, and 2) to simultaneously constrain the line-to-continuum time-delay using our formalism. The latter quantity is independently known from spectroscopic studies for most of the objects (Table 1), and may thus be used for benchmarking purposes.

\subsection{Individual Objects}

We first focus on instructive examples and then consider the reliability of our results as far as the statistical properties of the sample are concerned. Unless otherwise specified, we search for time-delays (i.e.,  peaks in the correlation functions) in the parameter space defined by $-300<\tau_c<300$\,days, $0<\tau_l<300$\,days, and  $0 \le \alpha \le 1$. 

\subsubsection{The "poster child": NGC\,5548}

\begin{figure}
\plotone{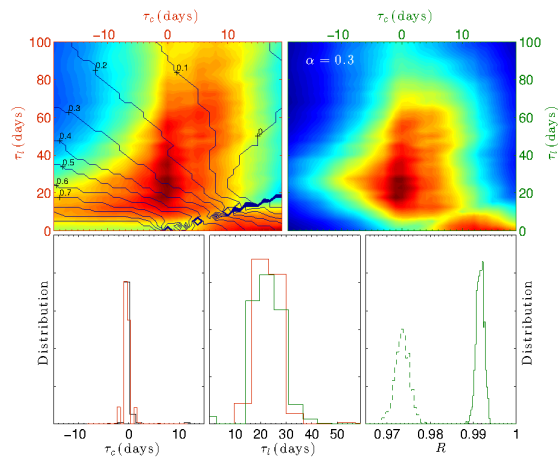}
\caption{2D cross-correlation analysis for NGC\,5548. Upper-left panel corresponds to the (projected) correlation function (no priors assumed) with $\alpha$ contours overlaid (redder colors correspond to higher correlation values). A peak is clearly visible at $\tau_c\sim 0$\,days and $\tau_l\sim 20$\,days. Right panel shows the correlation function restricted to the $\alpha=0.3$ plane. Lower panels show the $\tau_c$ and $\tau_l$ distributions with (green) and without (red) the use of priors on the value of $\alpha$, leading to consistent results. The signal is highly significant as can be inferred from the non overlapping $R$-distributions for the real (solid line) and permutated data (dashed lines).}
\label{N5548}
\end{figure}

NGC\,5548 is a well studied Seyfert 1 galaxy with a correlation function, $R(\tau_c,\tau_l,\alpha)$, peaking around $\tau_c\sim0$\,days, $\tau_l\sim 20$\,days, and at $\alpha\sim 0.3$. In particular, a single well-defined peak is visible in the computational domain (apart from reflection symmetry, as discussed in \S3). Using a spectroscopically-determined prior on $\alpha$ does not significantly alter the recovered time-delays. Further, the signal is highly significant according to our test, as can be seen from the non overlapping $R$-distributions using the real and permutated line templates, with the latter resulting in a maximal cross-correlation coefficient, $R$, being lower, by $\sim 2\%$ (recall that only the emission-line template is being permutated hence the detection of the emission line signal lurking in the $R$ light curve is significant). 

The deduced $\tau_l$ is in good agreement with spectroscopic RM results (Table 1). However, $\tau_c$ is smaller by an order of magnitude compared to $\tau_c^{\rm 1D}$, which leads us to conclude that, for NGC\,5548, the interband time delay is biased due to the contribution of non-continuum emission to the signal, hence its value has little to do with accretion  physics. As shall be further discussed below, the reported $\tau_c$ is also in better agreement with naive irradiated accretion disk models, and the spectroscopically-measured lag for an object of a similar luminosity \citep{col98}. 

\subsubsection{Marginal detection: 1E\,0754.6+392}

\begin{figure}
\plotone{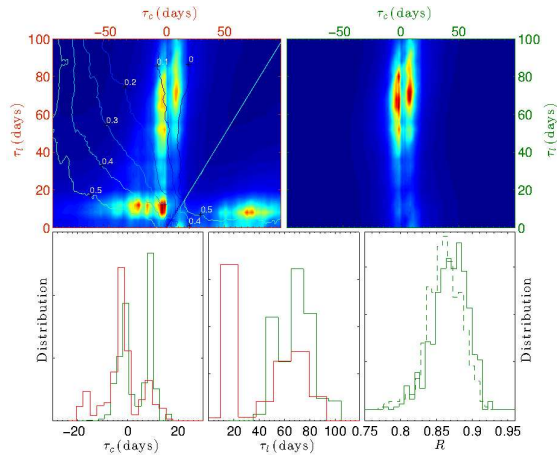}
\caption{Cross-correlation analysis for 1E\,0754.6+392 (see Fig. \ref{N5548} for the description of the different panels). Several peaks are visible in the upper-left panel (in addition to the expected reflection symmetry). When the solution is restricted to the $\alpha=0.1$ plane (upper-right panel), only the peak at $\tau_l\sim 70$\,days remains (see Table 1). This use of priors on $\alpha$ is able to suppress Lower panels show the $\tau_c,~\tau_l$ distributions with (green) and without (red) the use of priors on $\alpha$. Bottom-right panel shows the $R$-distributions obtained using our flux randomization for the real data (solid line) and the permutated data (dashed line). Clearly, the difference between the distributions is marginal at best (see text).}
\label{1E}
\end{figure}

1E\,0754.6+392 is a narrow line object, with a reported spectroscopic line-to-continuum time-delay of $\sim 100$\,days \citep{ser07}. Without the use of priors on the value of $\alpha$, two peaks are evident in the projected correlation function (Fig. \ref{1E}), around $\vert \tau_c \vert \sim \tau_l\sim$a few days, which correspond to values of $\alpha\sim 0.4$, and around $\tau_l\sim 70$\,days, $\tau_c\sim 0$\,days for $\alpha \sim 0.1$. While the former solution corresponds to a case in which the algorithm prefers  to reproduce $f_{lc}$ by ignoring the contribution of an emission line component to the $R$-band (see \S3.1), the latter solution is qualitatively consistent with the expected contribution of the H$\alpha$ emission line to the $R$-band, and with the spectroscopic time-delay. The bimodal nature of the solution is also manifested in the $\tau_c$ and $\tau_l$ distributions produced by our error estimation algorithm (Fig. \ref{1E}). By setting a prior of $\alpha=0.1$ (based on, e.g., single epoch spectroscopy; Table 1), it is possible to select for the physically-relevant solution wherein $\tau_c=8^{+5}_{-10}$\,days and $\tau_l=64^{+11}_{-17}$\,days. The latter value is in good agreement with the spectroscopic results of \citet[see our Table 1]{ser07}. 

Despite the favorable properties of our solution, $\alpha$ is only marginally greater than zero (Table 1), which implies that the emission line component is only marginally detected. Similarly, the significance of the solution, with no priors set (Fig. \ref{1E} and Table 1), is also marginal with $\lesssim 40\%$ chance of being spurious. To better understand this finding we note that (a) the signal to noise for this object is the lowest in our sample, and (b) although the peak in the correlation function occurs for $\alpha\sim 0.1$, some of the qualitative features of the correlation function are also evident when $\alpha \to 0$ (not shown here due to overall similarity with the right panel of Fig. \ref{1E}). This case merely reflects on the limitations of our interpolation scheme (see Appendix), and demonstrates that our significance scheme is able to capture such occurrences. Other computational schemes for evaluating the correlation function, such as those based on Gaussian processes \citep{rp92,pan11,zu11}, and more akin to TODCOR \citep{zm94} might lead to more robust results also in this case. Considering algorithms of this sort is beyond the scope of the present work. 

\begin{figure}
\plotone{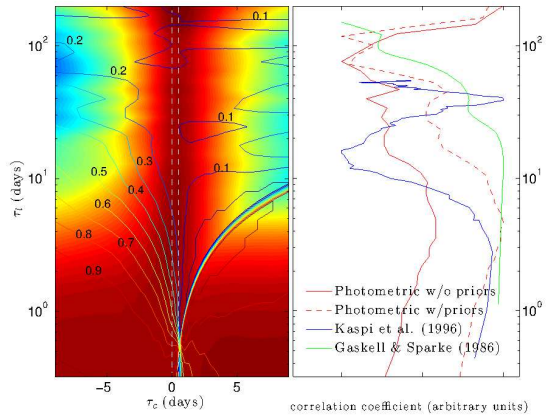}
\caption{Results for NGC\,4151. {\it Left:} $R$-statistics, with no priors, overlaid with $\alpha$ solution contours (note the logarithmically-scaled ordinate). The map shows two peaks: at $\tau_l\lesssim 10$\,days and beyond $\tau_l \sim 100$,days (both around $\tau_c\sim 0$\,days) wtihin the interval $0<\tau_c<0.5$\,days (marked by dashed white lines). {\it Right:} 1D projections of all the correlation functions (see legend) along the $\tau_l$ axis, and using the prior $\alpha=0.2$. Two well-separated peaks are evident: at $\tau_l\sim 10$\,days and at $\tau_l \lesssim 100$\,days, which are in qualitative agreement with the results of  spectroscopic RM technique \citep[green and blue lines, respectively]{gs86,kas96}. The first peak at $\lesssim 10$\,days is identified with the BLR time-delay for this object.}
\label{N4151}
\end{figure}

\subsubsection{Light curve de-trending: Akn\,120}

\begin{figure}
\plotone{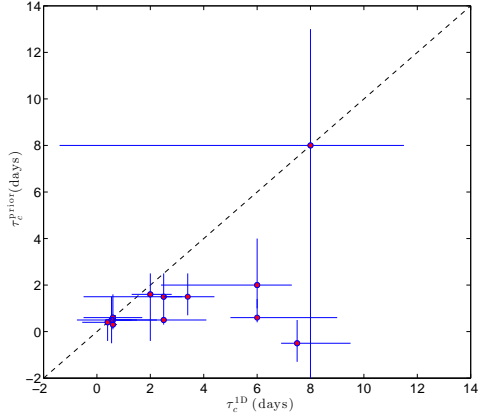}
\caption{Interband time-delays, $\tau_c^{\rm 1D}$ vs. continuum-to-continuum time-delays, $\tau_c$, for objects in the \citet{ser05} sample. Generally, $\tau_c^{\rm 1D}>\tau_c$, which results from the signal recovered by the cross-correlation function being contaminated by strong line emission in the $R$-band. In extreme cases, such as NGC\,5548 (see Table 1), $\tau_c^{\rm 1D}$ may exceed $\tau_c$ by more than an order of magnitude, thereby leading to considerably over-estimated accretion disk sizes. In contrast, the MCF formalism adopted here mitigates the effect of emission lines by explicitly accounting for their contribution to the signal.}
\label{comp1}
\end{figure}

The analyses of the light curves in this case, whether or not priors are assumed, yield statistically similar results, $\tau_l\sim\tau_l^{\rm prior}\simeq 7^{+5}_{-2}$ (not shown), which are significantly smaller than spectroscopic measured lag of $\sim 40$\,days  (Table 1). Inspection of the data in \citet[see their Fig. 1]{ser05} reveals significant variance at the lowest observable frequencies:  the mean flux during the first half of the campaign is significantly below its value toward the end. This leads to an effective non-stationary behavior of the light curves, which limits the usefulness of many methods for time-series analysis. 

A common practice to restore (quasi-) stationarity and obtain more reliable time-lag measurements, is to invoke de-trending \citep{wel99,cd12}. De-trending of the $V$ and $R$ light curves by a first degree polynomial, and repeating the analysis, we obtain a line-to-continuum delay, which is in better agreement with the spectroscopic results (Table 1). We note that de-trending appears to have a relatively small effect on the results for the other objects in our sample, yet all results quoted in Table 1 and discussed here were obtained using de-trended light curves.

\subsubsection{Multi-peak solutions: NGC\,4151}

Analyzing the data for this object we find evidence for a double peak structure in $R$, and whether or not priors are set on the value of $\alpha$ (Fig. \ref{N4151}). Specifically, there is a peak at $\tau_l\lesssim 10$\,days and a second peak at $\gtrsim 100$\,days. Both peaks lie along the stripe defined by $0<\tau_c<0.5$\,days (Table 1). A projection of the 2D correlation function along the $\tau_l$ axis (i.e., by averaging over its values along the $\tau_c$ dimension in a relevant $\tau_c$ interval; see Fig. \ref{N4151}) yields a 1D correlation function, which is comparable to the one obtained from a cross-correlation analysis of the spectroscopic data \citep[see also the right panel of Fig. \ref{N4151}]{kas96}. Specifically, a well separated two peak structure is evident, which roughly matches the structure seen in \citet{gs86,kas96}. In accordance with other works, we identify the first peak at $\lesssim 10$\,days with the size of the BLR in NGC\,4151, but note that a broader range of BLR scales may be applicable.

\subsection{Statistical Properties of the Sample}

In its current implementation, the method proposed here is able to detect a significant signal in most objects. Specifically, only $\sim30$\% of the sample are characterized by a signal whose significance is $<68$\%, with the confidence level for $\sim 30$\% of the objects being $\ge 90\%$  (see Table 1). The least significant signal detections are also characterized $\alpha$-values that are marginally consistent with zero, and their lags deviate the most from the spectroscopic lags (Table 1 and Fig. \ref{comp1}). The light curves of AGN for which emission line signals were not securely detected are broadly characterized by a combination of a lower reduced variability measure  (e.g., the case of 1E\,0754.6+392 having the largest photometric errors in the sample), and/or having a smaller number of photometric points. 

Quite generally, we find that $\tau_c < \tau_c^{\rm 1D}$, in some cases by as much as an order of magnitude (Fig \ref{comp1} and Table 1, and note the case of NGC\,5548), and attribute that to the presence of a relatively strong emission line contribution to the $R$-band, which biases the cross-correlation function (\S2). Considering the sample as a whole, we find a mean $\tau_c^{\rm 1D}/\tau_c \sim 3$, implying that interband time-delays are poor tracers of $\tau_c$, and that their use may lead to erroneous results concerning the sizes of accretion disks.

\begin{figure}
\plotone{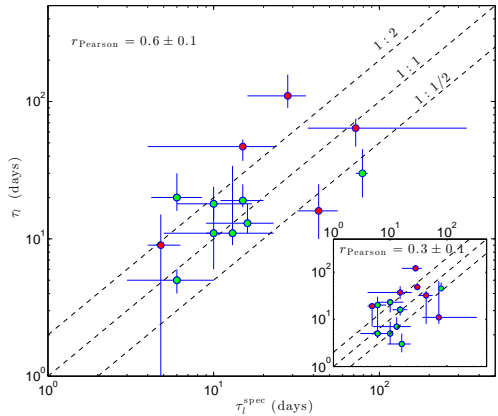}
\caption{Line-to-continuum time-delays, $\tau_l$, vs. published, spectroscopically determined, Balmer emission line lags, $\tau_l^{\rm spec}$, for objects in the \citet{ser05} sample. Good agreement is obtained between the photometric and the spectroscopic results. Specifically, the use of priors on $\alpha$ leads to a tighter correlation between $\tau_l$ and $\tau_l^{\rm spec}$ (the inset shows the photometric lags which were deduced without the use of priors). Pearson's $r$-values quoted in each panel were obtained for logarithmically-scaled data. Green data points correspond to results whose significance is $>68$\% (Table 1). Most of the scatter between the spectroscopic and the photometric lags (using priors) may be attributed to the non-contemporaneous nature of the spectroscopic and the photometric campaigns, and given the fact that BLR sizes, as deduced from RM of the Balmer lines, are known to vary with time (see text).}
\label{comp2}
\end{figure}

We find good agreement between $\tau_l$ and $\tau_l^{\rm spec}$ (Table 1 and Fig. \ref{comp2}), and that the two measurements are more tightly correlated when priors on $\alpha$ are incorporated into the analysis (leading to Pearson's $r\sim 0.6$; see Fig. \ref{comp2}). While some of the scatter may be attributed to those objects with less secure lag measurements, some residual scatter remains even for the best-case examples (Table 1). A potentially relevant source for the residual scatter is the time-varying nature of the BLR size in AGN as traced by line emission \citep{pet04}. In particular, a range of lags, spanning a factor $\sim 4$ has been measured for the H$\beta$ line in NGC\,5548 at different epochs, which, if characteristic of AGN [see also Fig. \ref{N4151} where the results of \citet{gs86} and \citet{kas96} differ by a similar factor], could fully account for the observed scatter.

Comparing columns (7) and (10) in table 1, we find a hint for $\alpha \gtrsim \alpha^i$. A proper investigation into these subtle effects is currently unwarranted yet we note that this might reflect on additional emission components, other than H$\alpha$, which are being emitted on BLR-scales and contribute to the $R$-band signal, such as Paschen recombination emission. Alternatively, it may reflect on the relative contribution of emission lines to the varying component in the $R$-band being larger than their relative contribution to the flux.

Attempting to quantify possible systematic effects between the photometric lags and the spectroscopic ones is currently unwarranted due to small number statistics. Nevertheless, there is no clear indication for a bias, which suggests that the method is immune to a small contribution of emission line signal to the $V$-band. This conclusion is also in accordance with the findings of \citet{cd11}, \citet{cd12}, \citet{ed12}, and \citet{po13}.  Care should be taken, however, when interpreting the results in cases where the broadband data contain comparable contributions from several emission lines (or other emission components) with very different time delays. A possible means to treat such cases may involve the generalization of our model to higher dimensions, yet such a treatment is beyond the scope of the present work, and is not warranted by the current data.

\section{Implications for the Study of AGN}

Although the sample is small, and of no match to what will be possible to achieve with future surveys, it is nevertheless worth placing our results in a physical context. 

\subsection{The Photometric $R_{\rm BLR}-L$ Relation}

In \citet{cd11}, the concept of broadband photometric RM was introduced as a means for (statistically) estimating the size of the BLR to potentially unprecedented precision using high-quality data for many objects in the era of large photometric surveys, such as LSST. In particular, the relatively small number of visits per filter in the sample of quasars used in that work yielded significant delays in only a handful of objects, and the advantage of the method was demonstrated mainly on statistical grounds. Better datasets demonstrated that such an approach may be used to determine the BLR size also in individual objects \citep{ed12,po13}.

\begin{figure*}
\plottwo{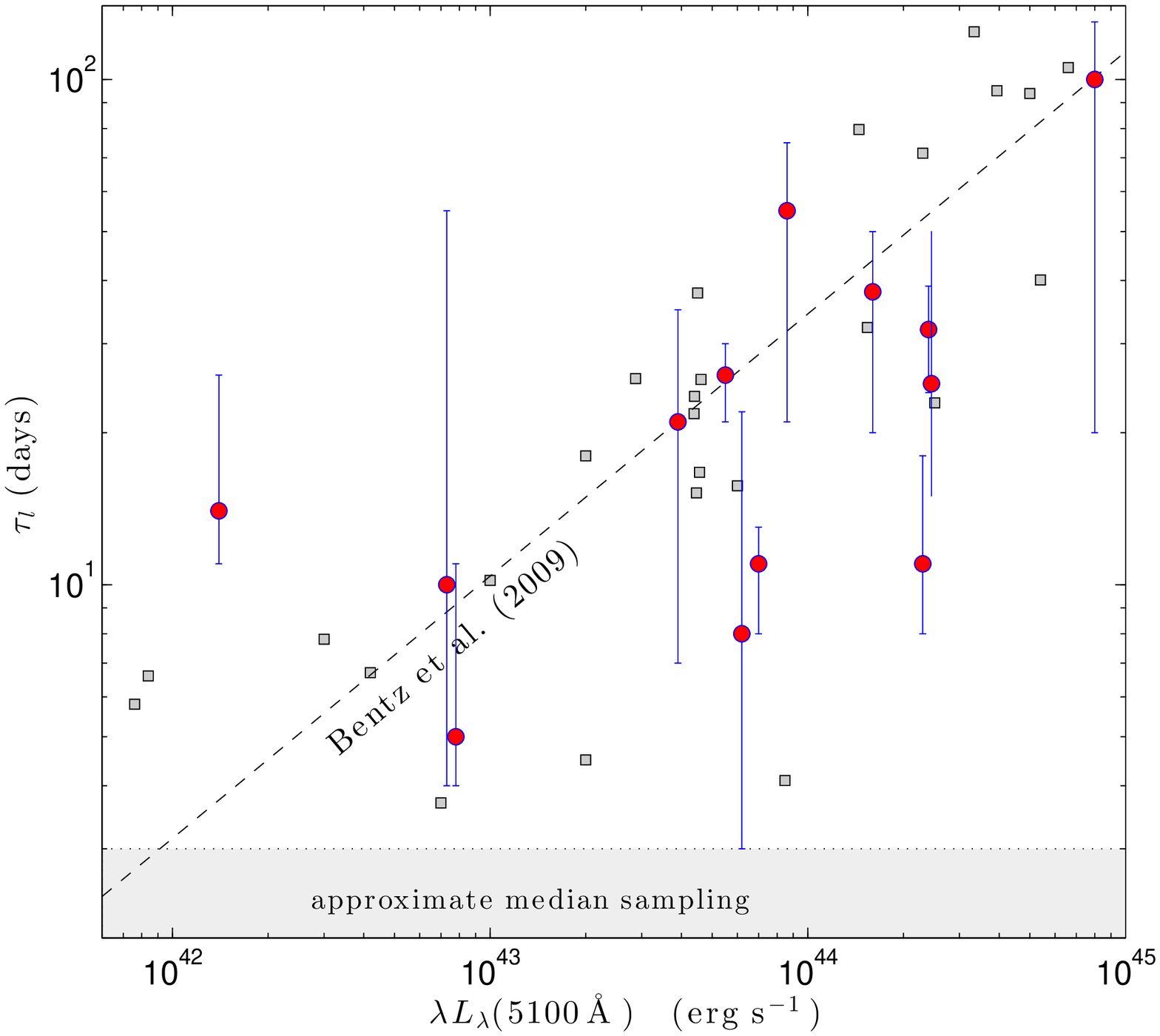}{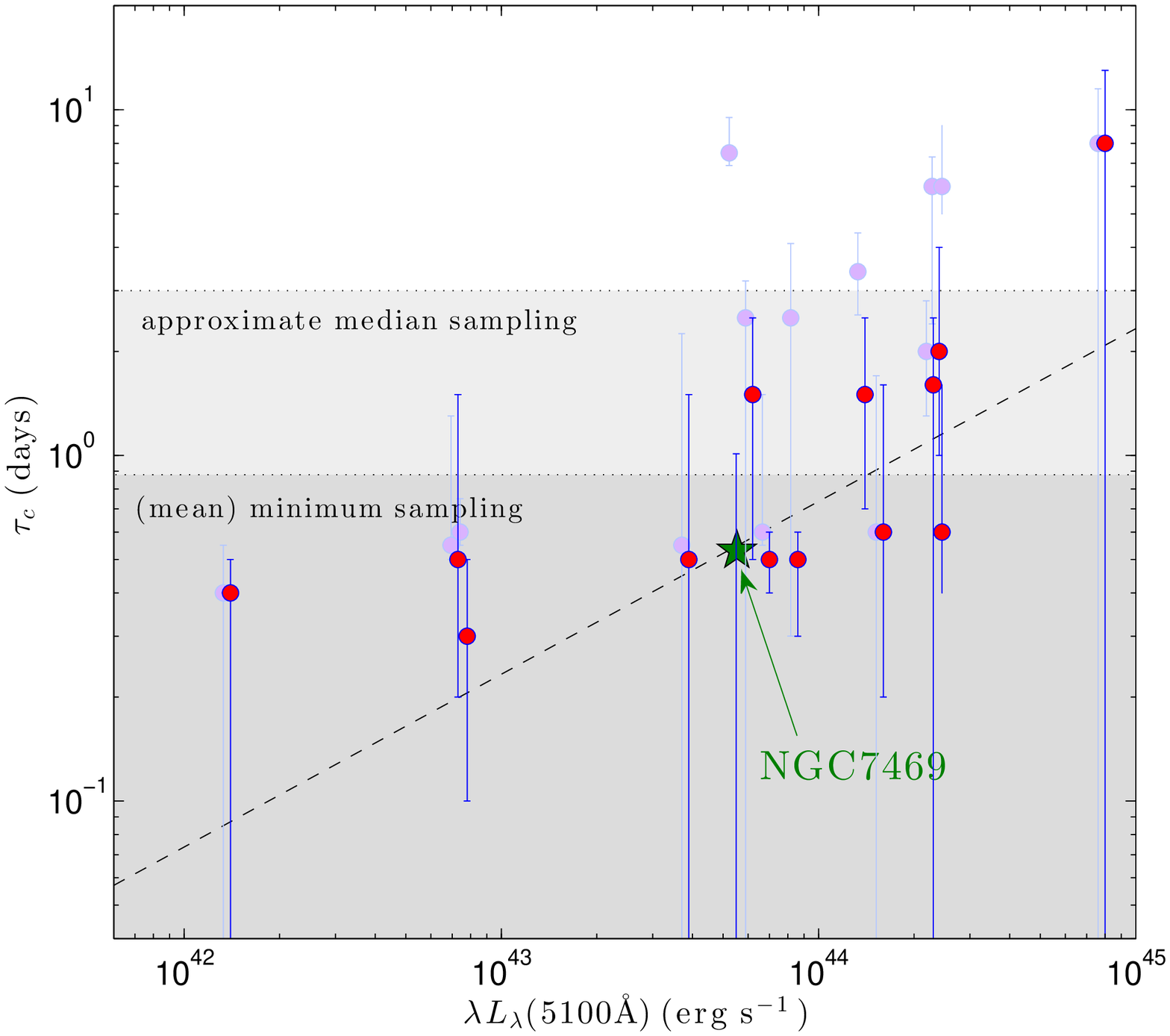}
\caption{Lag-luminosity relations for AGN. {\it Left:} The line-to-continuum time-delay, $\tau_l$ vs. the optical luminosity relation (results obtained using priors on $\alpha$ are shown) is broadly consistent with the spectroscopic relation of \citet{ben09}. Specifically, the scatter about the powerlaw is consistent for the photometric and the spectroscopic (gray points) samples. 
{\it Right:} continuum-to-continuum time-delays, $\tau_c$, as measured from the broadband photometric data of \citet{ser05}. Results are qualitatively consistent with lying on a powerlaw extending $\sim 3$ decades in luminosity with $\tau_c\sim [\lambda L_\lambda (5100\,{\rm \AA})]^{1/2}$ (dashed line using the spectroscopic time delay of NGC\,7469 as pivot). In particular, our photometric time-lag measurements are consistent with the spectroscopic result of \citet{col98} for NGC\,7469 for similar luminosity objects. As shown, interband time-delays (Table 1) are biased to larger values (see light-shaded data points and note that the luminosity was slightly downshifted for clarity).}
\label{lagl}
\end{figure*}

Here, by adopting a refined RM approach with a revised significance and error estimation algorithm, and working with superior photometric data, we are able to securely measure time-delays associated with the BLR in most objects, and are therefore able, for the first time, to plot the photometric version of the size-luminosity relation for the BLR in Fig. \ref{lagl}\footnote{Redshift effects are negligible for this sample of AGN hence ignored.}. Evidently, the photometric relation statistically traces the spectroscopic one, and both have qualitatively similar scatter around the \citet{ben09} relation.  Clearly, with high-quality data in several bands, for numerous objects, much more precise versions of figure \ref{lagl} may be obtained, also for different sub-classes of AGN, and for different emission lines. 

\subsection{The Irradiated Accretion Disk Model}

While standard accretion disk models \citep{ss73} have been very successful in accounting for some properties of AGN emission, many uncertainties remain, and the search is on for additional reliable means to probe their physics. Specifically, a fundamental open question, in the context of AGN accretion, concerns the size of the region from which the bulk of the optical emission is emitted. A promising route to shed light on the size of those spatially-unresolved regions is via RM \citep[and references therein]{col98,ser05,cak07}. 

For standard accretion disks characteristic of most AGN, the predicted optical emission is of a powerlaw form dependence on photon energy, and its luminosity, $\lambda L_\lambda (5100\,{\rm \AA}) \sim (M_{\rm BH}\dot{M})^{2/3}$ \citep[where $M_{\rm BH}$ is the black hole mass, and $\dot{M}$ the accretion rate]{bec87}. On the other hand, the effective accretion disk size, at a given restframe wavelength, scales as $(M_{\rm BH}\dot{M})^{1/3}$ \citep{col98}. Therefore, the crossing time for perturbations over a part of the optical disk, $\tau_c\sim [\lambda L_\lambda (5100\,{\rm \AA})]^{1/2}$. 

Our lag-luminosity relation is shown in figure \ref{lagl}, with the value corresponding to the spectroscopic measurement of NGC\,7469  over-plotted \citep[converted to the wavelength range covered here; see their Fig. 7]{col98}. Evidently, the continuum lags deduced here are of the right order of magnitude for objects with a similar luminosity to NGC\,7469. Specifically, for objects with luminosities $\pm0.5$\,dex of NGC\,7469, the median $\tau_c$ agrees with the spectroscopically measured lag in NGC\,7469, while most interband time-delays in this luminosity range are greater than this value by a factor $\sim 3$ (in the latter case the standard deviation of the lags is also larger by a factor $\sim 4$).  Furthermore, the lag-luminosity relation is qualitatively consistent with the expected power-law behavior from a thin irradiated accretion disk model, using NGC\,7469's time-delay as a pivot (note that results at the low-luminosity end may be affected by under-sampling). We do not, however, attempt to provide more quantitative statements in this work, and refer the reader to \citet{c12} where a more refined analysis, based on multi-band data is carried out.

\section{Summary}

Reverberation mapping (RM) has proven to be a valuable technique for studying spatially unresolved regions in AGN. Specifically, it has been implemented, using spectroscopic data, to measure the size of the BLR in $\sim 50$ AGN, and has also been used to place constraints on the size of the accretion disk in a few systems. Nevertheless, spectroscopic datasets that are adequate for RM are scarce. In contrast, photometric data are relatively easy to acquire yet disentangling the various emission processes that contribute to the signal is not straightforward, which could lead to erroneous conclusions.

Motivated by upcoming (photometric) surveys that will provide exquisite light curves for numerous AGN, we propose to generalize the cross-correlation scheme and work instead with a multivariate correlation function (MCF). The advantages of the proposed approach are the following:
\begin{enumerate}[(i)]
\item
It can simultaneously constrain continuum-to-continuum (accretion disk), and line-to-continuum (BLR) time delays, as well as the relative contribution of their respective processes to the signal. As such, it improves upon current cross-correlation techniques for lag determination.
\item
The method is equally applicable to photometric (broad- and narrow-band) and spectroscopic data. As such, it allows for the reliable determination of the time-delays even in cases where spectral decomposition of line and continuum processes is challenging.
\item
Prior knowledge of one or more of the variables in the problem can be easily incorporated into the analysis thereby leading to more robust constraints on the remaining model parameters. 
\end{enumerate}

Applying the method to the high-quality broadband photometric data for 14 AGN in the \citet{ser05} sample, we are able to simultaneously determine accretion disk scales and BLR sizes in those sources. In particular, our photometric line-to-continuum time-delays for individual objects are in good agreement with spectroscopic Balmer line lag measurements. This further demonstrates (see also \citet{cd11}) that, provided photometric data are adequate, spectroscopic data are not a prerequisite for BLR size determination. In addition, we provide the first {\it reliable} accretion disk scale vs. AGN luminosity relation, which is in qualitative agreement with theoretical expectations from standard irradiated accretion disk models.

\acknowledgements 

We are grateful to H. Netzer for thought-provoking discussions that triggered our interest in the problem, and to A. Laor, E. Behar, E. Daniel, S. Kaspi, S. Rafter, N. Scoville, and E. Ribak for many fruitful conversations and good advice. We thank the referee for  valuable comments. This research has been supported in part by a FP7/IRG PIRG-GA-2009-256434 as well as by grant 927/11 from the Israeli Science Foundation and the Jack Adler Foundation awarded to D.C. S.Z. acknowledges partial support from the DFG via German-Israeli Project Cooperation grant STE1869/1-1.GE625/15-1 as well as partial support from the Israeli Ministry of Science and Technology via grant 3-9082.



\appendix

\section{Evaluation of the MCF}

The Pearson correlation coefficient, as a function of the continuum time-delay, $\tau_c$, the line to continuum time-delay, $\tau_l$, and the contribution of the emission line to the flux in the line-rich band, $\alpha$, is of the general form:
\begin{equation}
R(\tau_c,\tau_l,\alpha)=\frac{\sum_i [f_{lc}^{\tau_c,\tau_l}(t_i)-\bar{f}_{lc}^{\tau_c,\tau_l}] [f^m_{lc}(t_i;\tau_c,\tau_l) -\bar{f}^m_{lc}(\tau_c,\tau_l)]}{ \left \{ \sum_i [f_{lc}^{\tau_c,\tau_l}(t_i)-\bar{f}_{lc}^{\tau_c,\tau_l}]^2 \right \}^{1/2} \left \{ \sum_i [f^m_{lc}(t_i;\tau_c,\tau_l)-\bar{f}^m_{lc}(\tau_c,\tau_l)]^2 \right \}^{1/2} }.
\label{r}
\end{equation}
A linear interpolation scheme is used to calculate $R$ for any choice of $\tau_c,\tau_l$. Specifically, three time series are involved, two of which are identical but are time-shifted versions of each other (recall the definition of $f_{lc}^m$ in Eq. \ref{flcp}): $f_{lc}(t_i),~f_c(t_j-\tau_c),~f_c(t_j-\tau_l)$, where $i \in [1,2,3,....N]$ and $j \in [1,2,3... M]$, where $N,~M$ are the number of visits in $f_{lc},~f_c$, respectively. A new time-series vector is then formed $t_k \in \{ \{ t_1,t_2,t_3...t_N \}, \{ t_1-\tau_c,t_2-\tau_c,t_3+\tau_c...t_M-\tau_c \},  \{ t_1-\tau_l,t_2-\tau_l,t_3-\tau_l...t_M-\tau_l \} \}$, which is sorted, and repeating time-stamps discarded. Further, only those times, $t_k$ which do not require extrapolation of any of the (shifted) light curves are kept. This time series is then used to calculate $R(\tau_c,\tau_l,\alpha)$, by interpolating on all light curves at $t_k$, as required. For this reason, the particular $f_{lc}$ that enters equation \ref{r}, implicitly depends on $\tau_c$ and $\tau_l$, and is therefore denoted as $f_{lc}^{\tau_c,\tau_l}$.  The interpolation scheme used here is, essentially, the symmetrized partial-interpolation method often used in RM studies \citep{pet93}. Experimenting with other methods of interpolation and evaluation of equation \ref{ccf} \citep[whose TODCOR algorithm operates in Fourier space]{zm94} may be of interest yet are beyond the scope of the present work.



\begin{thebibliography}{}

\bibitem[Bachev(2009)]{bac09} Bachev, R.~S.\ 2009, \aap, 493, 907

\bibitem[Bechtold et al.(1987)]{bec87} Bechtold, J., Czerny, B., Elvis, M., Fabbiano, G., \& Green, R.~F.\ 1987, \apj, 314, 699

\bibitem[Bentz et al.(2006)]{ben06} Bentz, M.~C., Peterson, B.~M., Pogge, R.~W., Vestergaard, M., \& Onken, C.~A.\ 2006, \apj, 644, 133

\bibitem[Bentz et al.(2009)]{ben09} Bentz, M.~C., Peterson, B.~M., Netzer, H., Pogge, R.~W., \& Vestergaard, M.\ 2009, \apj, 697, 160

\bibitem[Breedt et al.(2009)]{bre09} Breedt, E., Ar{\'e}valo, P., McHardy, I.~M., et al.\ 2009, \mnras, 394, 427

\bibitem[Cackett et al.(2007)]{cak07} Cackett, E.~M., Horne, K., \& Winkler, H.\ 2007, \mnras, 380, 669 

\bibitem[Chelouche (2013)]{c12} Chelouche, D.,\ 2013, \apj, submitted

\bibitem[Chelouche \& Daniel(2012)]{cd11} Chelouche, D., \& Daniel E.\ 2012, \apj, 747, 62 

\bibitem[Chelouche et al.(2012)]{cd12} Chelouche, D., Daniel, E., \& Kaspi, S.\ 2012, \apjl, 750, L43

\bibitem[Collier et al.(1998)]{col98} Collier, S.~J., Horne, K., Kaspi, S., et al.\ 1998, \apj, 500, 162

\bibitem[Collier(2001)]{col01} Collier, S.\ 2001, \mnras, 
325, 1527

\bibitem[Denney et al.(2010)]{den10} Denney, K.~D., Peterson, B.~M., Pogge, R.~W., et al.\ 2010, \apj, 721, 715

\bibitem[Doroshenko et al.(2005)]{dor05} Doroshenko, V.~T., Sergeev, S.~G., Merkulova, N.~I., Sergeeva, E.~A., \& Golubinsky, Y.~V.\ 2005, \aap, 437, 87

\bibitem[Edri et al.(2012)]{ed12} Edri, H., Rafter, S.~E., Chelouche, D., Kaspi, S., \& Behar, E.\ 2012, \apj, 756, 73

\bibitem[Gaskell(2007)]{gas07} Gaskell, C.~M.\ 2007, The Central Engine of Active Galactic Nuclei, 373, 596

\bibitem[Gaskell \& Sparke(1986)]{gs86} Gaskell, C.~M., \& Sparke, L.~S.\ 1986, \apj, 305, 175

\bibitem[Giveon et al.(1999)]{giv99} Giveon, U., Maoz, D., Kaspi, S., Netzer, H., \& Smith, P.~S.\ 1999, \mnras, 306, 637

\bibitem[Grier et al.(2012)]{gr12} Grier, C.~J., Peterson, B.~M., Pogge, R.~W., et al.\ 2012, \apj, 755, 60

\bibitem[Haas et al.(2011)]{ha11} Haas, M., Chini, R., Ramolla, M., et al.\ 2011, \aap, 535, A73

\bibitem[Kaspi et al.(1996)]{kas96} Kaspi, S., Maoz, D., Netzer, H., et al.\ 1996, \apj, 470, 336

\bibitem[Kaspi et al.(2000)]{kas00} Kaspi, S., Smith, P.~S., Netzer, H., Maoz, D., Jannuzi, B.~T., \& Giveon, U.\ 2000, \apj, 533, 631

\bibitem[Kaspi et al.(2005)]{kas05} Kaspi, S., Maoz, D., Netzer, H., et al.\ 2005, \apj, 629, 61

\bibitem[Koptelova \& Oknyanskij(2010)]{kop10} Koptelova, E., \& Oknyanskij, V.\ 2010, The Open Astronomy Journal, 3, 184

\bibitem[Maoz et al.(1991)]{mao91} Maoz, D., Netzer, H., Mazeh, T., et al.\ 1991, \apj, 367, 493

\bibitem[Onken et al.(2004)]{on04} Onken, C.~A., Ferrarese, L., Merritt, D., et al.\ 2004, \apj, 615, 645

\bibitem[Pancoast et al.(2011)]{pan11} Pancoast, A., Brewer, B.~J., \& Treu, T.\ 2011, \apj, 730, 139

\bibitem[Peterson(1993)]{pet93} Peterson, B.~M.\ 1993, \pasp, 105, 247

\bibitem[Peterson et al.(1998)]{pet98a} Peterson, B.~M., Wanders, I., Bertram, R., et al.\ 1998, \apj, 501, 82

\bibitem[Peterson et al.(1998)]{pet98} Peterson, B.~M., Wanders, I., Horne, K., et al.\ 1998, \pasp, 110, 660

\bibitem[Peterson et al.(2004)]{pet04} Peterson, B.~M., et al.\ 2004, \apj, 613, 682

\bibitem[Pozo Nu{\~n}ez et al.(2012)]{po12} Pozo Nu{\~n}ez, F., Ramolla, M., Westhues, C., et al.\ 2012, \aap, 545, A84

\bibitem[Pozo Nu{\~n}ez et al.(2013)]{po13} Pozo Nu{\~n}ez, F., Westhues, C., Ramolla, M., et al.\ 2013, \aap, in press

\bibitem[Rafter et al.(2013)]{raf13} Rafter S. E., Kaspi, S., Chelouche D., et al.\ 2013, \apj, submitted

\bibitem[Rybicki \& Press(1992)]{rp92} Rybicki, G.~B., \& Press, W.~H.\ 1992, \apj, 398, 169

\bibitem[Salamanca et al.(1994)]{sal94} Salamanca, I., Alloin, D., Baribaud, T., et al.\ 1994, \aap, 282, 742

\bibitem[Salpeter(1964)]{sal64} Salpeter, E.~E.\ 1964, \apj, 140, 796

\bibitem[Sergeev et al.(2005)]{ser05} Sergeev, S.~G., Doroshenko, V.~T., Golubinskiy, Y.~V., Merkulova, N.~I.,  \& Sergeeva, E.~A.\ 2005, \apj, 622, 129

\bibitem[Sergeev et al.(2007)]{ser07} Sergeev, S.~G., Klimanov, S.~A., Chesnok, N.~G., 
\& Pronik, V.~I.\ 2007, Astronomy Letters, 33, 429 

\bibitem[Shakura \& Sunyaev(1973)]{ss73} Shakura, N.~I., \& Sunyaev, R.~A.\ 1973, \aap, 24, 337

\bibitem[Wanders et al.(1997)]{wan97} Wanders, I., Peterson, B.~M., Alloin, D., et al.\ 1997, \apjs, 113, 69

\bibitem[Welsh(1999)]{wel99} Welsh, W.~F.\ 1999, \pasp, 111, 1347

\bibitem[Woo et al.(2010)]{woo10} Woo, J.-H., Treu, T., Barth, A.~J., et al.\ 2010, \apj, 716, 269

\bibitem[Zel'dovich(1964)]{zel64} Zel'dovich, Y.~B.\ 1964, Soviet Physics Doklady, 9, 195

\bibitem[Zu et al.(2011)]{zu11} Zu, Y., Kochanek, C.~S., \& Peterson, B.~M.\ 2011, \apj, 735, 80

\bibitem[Zucker \& Mazeh(1994)]{zm94} Zucker, S., \& Mazeh, T.\ 1994, \apj, 420, 806


\end{thebibliography}
\end{document}